\RequirePackage{fix-cm}
\documentclass[twocolumn,epjc3,final]{svjour3}  
\smartqed  % flush right qed marks, e.g. at end of proof
 \usepackage{mathrsfs}
\usepackage{amsmath, txfonts}
\usepackage{graphicx,bm,booktabs,bm}
\usepackage{longtable,lscape}
\usepackage{ulem} 
\usepackage{overpic,multirow,color}
\definecolor{cover}{rgb}{0.77,0.87,0.88}
\definecolor{blueone}{rgb}{0.1,0.1,.7}
\definecolor{citec}{rgb}{0.14,0.47,0.09}
\definecolor{two}{rgb}{0.0,0.5,0.}
\definecolor{three}{rgb}{.5,.1,0.15}
\usepackage[bookmarks=true,bookmarksopen=false,plainpages=false,breaklinks=true,
   bookmarksnumbered=true,hypertexnames=false,
   filecolor=blue,urlcolor=three,menucolor=three,
   linkcolor=three,citecolor=blueone, colorlinks,
   anchorcolor=blue,runcolor=pink,frenchlinks=red
   pdfstartview=FitH,pdftitle=title,%
   pdfauthor=author]{hyperref}
              \allowdisplaybreaks[4]

\graphicspath{{Figures/}} %

\journalname{Eur. Phys. J. C}
\begin{document}
\title{$X_0(2900)$ and $\chi_{c0}(3930)$ in  process $B^+\to D^+ D^- K^+$}
\author{Zuo-Ming Ding, Qi Huang \and Jun He\thanksref{e1}
}                     % Do not remove
\thankstext{e1}{Corresponding author: junhe@njnu.edu.cn}
\institute{Department of  Physics and Institute of Theoretical Physics, Nanjing Normal University,
Nanjing 210097, China}

\date{Received: date / Revised version: date}
\maketitle

\abstract{

This study investigates the nature of the $X_0(2900)$ and $\chi_{c0}(3930)$
based on experimental results of the process $B^+$$\to$$D^+ D^- K^+$. We focus on
the S-wave $D^{*-}K^{*+}$ and $D_s^+D_s^-$ molecular states, which can be
related to the $X_0(2900)$ and $\chi_{c0}(3930)$, respectively. Using effective Lagrangians,
we construct the potential kernel of the $D^{*-}K^{*+}$$-$$D^{-}K^{+}$ and
$D_s^+D_s^-$$-$$D^+D^-$ interactions with a one-boson-exchange model, and
determine the scattering amplitudes and their poles through a quasipotential
Bethe-Salpeter equation approach. By incorporating the potential kernel into the
three-body decay process $B^+$$\to$$D^+ D^- K^+$, we evaluate the $D^-K^+$ and
$D^+D^-$ invariant mass spectra, as well as the Dalitz plot, using Monte
Carlo simulation. A satisfactory fit to the $D^-K^+$ and $D^+D^-$ invariant mass
spectra is achieved after introducing additional Breit-Wigner resonances,
 $X_1(2900)$, $\psi(3770)$, and $\chi_{c2}(3930)$.
Prominent signals of the $X_0(2900)$ and $\chi_{c0}(3930)$ states appear  as
peaks in the $D^-K^+$ and $D^+D^-$ invariant mass spectra near 2900 and 3930
MeV, respectively. Clear event concentration from the $X_0(2900)$ and
$\chi_0(3930)$ is evident as strips in the Dalitz plot. The results
suggest that both $X_0(2900)$ and $\chi_{c0}(3930)$ can be interpreted as molecular
states, with the inclusion of $X_1(2900)$ and $\chi_2(3930)$ necessary to
describe structures in the $D^-K^+$ and $D^+D^-$ invariant mass
spectra, respectively.

 } %end of abstract

\section{Introduction}\label{sec1}

In 2020, the LHCb Collaboration observed two pairs of states, $X_{0,1}(2900)$
and $\chi_{c0,2}(3930)$, in the $D^-K^+$ and $D^+D^-$ invariant mass
distributions of the $B^+$$\to$$D^+ D^- K^+$ decay. The masses and widths of these
states were determined by LHCb and are listed in Table
\ref{LHCbdata}~\cite{LHCb:2020bls,LHCb:2020pxc}.

\renewcommand\tabcolsep{0.32cm}
\renewcommand{\arraystretch}{1.2}
\begin{table}[!tb]
\centering
\caption{Masses and widths for the $\chi_{c0,2}(3930)$ and $X_{0,1}(2900)$ resonances determined in Refs.~\cite{LHCb:2020bls,LHCb:2020pxc}
\label{LHCbdata}}
%%%%%%%%%%%%
\begin{tabular}{ccc }
\hline
Resonance & Mass (GeV) & Width (MeV) \\
\hline
$\chi_{c0}(3930)$ & 3.9238 $\pm$ 0.0015 $\pm$ 0.0004 & 17.4 $\pm$ 5.1 $\pm$ 0.8 \\
$\chi_{c2}(3930)$ & 3.9268 $\pm$ 0.0024 $\pm$ 0.0008 & 34.2 $\pm$ 6.6 $\pm$ 1.1 \\
$X_0(2900)$ & 2.866$ \pm$ 0.007 $\pm$ 0.002 & 57 $\pm$ 12 $\pm$ 4 \\
$X_1(2900)$ & 2.904 $\pm$ 0.005 $\pm$ 0.001 & 110 $\pm$ 11 $\pm$ 4 \\
\hline
\end{tabular}
%%%%%%%%%%%%
\end{table}

Among these discovered structures, $X_{0}(2900)$, also named
$T^{\bar{c}\bar{s}0}(2870)^0$~\cite{LHCb:2024vfz}, is the first exotic candidate
with four different flavors, attracting significant attention. Despite various
theoretical interpretations regarding the nature of $X_{0}(2900)$, such as the
compact tetraquark~\cite{Zhang:2020oze,Wang:2020xyc,He:2020jna,Wang:2020prk} or
triangle singularity~\cite{Liu:2020orv,Burns:2020epm}, the measured mass of
$X_{0}(2900)$ is close to the $\bar{D}^{*}K^{*}$ threshold, making the molecular
picture appealing.
In this context, the $X_{0}(2900)$ state, as well as the related
$\bar{D}^{(*)}K^{(*)}$ system, have been investigated within different frameworks,
such as chiral effective field
theory~\cite{Molina:2020hde}, QCD sum rule
methods~\cite{Chen:2020aos,Agaev:2020nrc,Mutuk:2020igv}, one-boson-exchange
model~\cite{Liu:2020nil}, and effective Lagrangian approach~\cite{Xiao:2020ltm},
to verify the molecular structure of $X_{0}(2900)$ and explore other possible
molecules.
In our previous work~\cite{He:2020btl}, we studied the
$D^{*-}K^{*+}$ interaction using a quasipotential Bethe-Salpeter equation (qBSE)
approach and constructed a one-boson-exchange potential with the help of heavy
quark and chiral symmetries. An improved method was proposed in another of our
previous works~\cite{Kong:2021ohg}, where we adopted hidden-gauge Lagrangians to
construct the potential kernels, making the theoretical framework more
self-consistent. The results of both studies suggest that $X_0(2900)$ can be
explained as a $D^{*-}K^{*+}$ molecular state with $I(J^P)=0(0^+)$.

In contrast to the $X_0(2900)$ state, the $\chi_{c0}(3930)$ state has received
relatively little attention until the discovery of the $X(3960)$ state in the
$D_s^+D_s^-$ invariant mass spectra~\cite{LHCb:2022aki}. These two states
have similar masses and widths, as well as the preferred $J^{PC} = 0^{++}$.
Consequently, many investigations have proposed coupled-channel analyses to
uncover the nature of the $\chi_{c0}(3930)$/$X(3960)$. It is common
to interpret $\chi_{c0}(3930)$/$X(3960)$ as $D_s^+D_s^-$$-$$D^+D^-$
molecules~\cite{Bayar:2022dqa,Chen:2023eix}, given that the mass
of the $X(3960)$ state is close to the $D_s^+D_s^-$ threshold.
In our previous work, we employed the three-body decay $B^+$$\to$$(D_s^+D_s^-
/D^+D^-)K^+$ in a qBSE approach to investigate the $\chi_{c0}(3930)$/$X(3960)$
resonance structure observed by LHCb, assuming that the
$\chi_{c0}(3930)$/$X(3960)$ is S-wave $D_s^+D_s^-$
molecule~\cite{Ding:2023yuo}. Our results indicated that the $X(3960)$ state
can be well reproduced in the $D_s^+D_s^-$ invariant mass spectrum, and the
$\chi_{c0}(3930)$ state can be observed in the $D^+D^-$ invariant mass spectrum,
albeit with a very small width.
As our previous model~\cite{Ding:2023yuo} only considered the $\chi_{c0}(3930)$,
and the experimental resonances in the invariant mass spectrum  were not perfectly reproduced, it is plausible to
suggest that the $\chi_{c2}(3930)$, which is also located near 3930 MeV but with
$J=2$, may significantly contribute to the structure observed near 3930 MeV in
the $D^+D^-$ invariant mass spectrum. Further investigations are needed to
substantiate this hypothesis and to understand the internal configurations of
the $\chi_{cJ}$ states.

Many theoretical studies have attempted to elucidate the nature of $X_0(2900)$
and $\chi_{c0}(3930)$ using diverse methodologies and experimental inputs.
However, these two states are often  investigated separately. Despite the
significance of resonance structures in both $D_s\bar{D}_s$ and
charm-strange systems, only a few amplitude analyses of the decay process 
$B^+$$\to$$D^+ D^- K^+$ have considered the combined contributions of both 
$X_0(2900)$ and $\chi_{c0}(3930)$. In the present research, we aim to
investigate the $D^-K^+$ and $D^+D^-$ invariant mass spectra and the Dalitz
plot of the $B^+\to D^+ D^- K^+$ process, taking into account the rescatterings
associated with the molecular states corresponding to $X_0(2900)$ and
$\chi_{c0}(3930)$. Through a comparison with experimental data from LHCb, we
will delve into the mechanism of the $B^+\to D^+ D^- K^+$ process and discuss
the nature of $X_0(2900)$ and $\chi_{c0}(3930)$ in this context.

In the following section, we will outline the theoretical framework utilized to
investigate the $B^+$$\to$$D^+ D^- K^+$process. We will provide a comprehensive
explanation of the mechanism and Lagrangians involved in
this process through the intermediate states $X_0(2900)$ and $\chi_0(3930)$. The
potential kernels for each process will be established and incorporated into the
qBSE to calculate the invariant mass
spectra and Dalitz plot.  In section~\ref{sec3}, we will analyze the
invariant mass spectra of the $B^+$$\to$$D^+ D^- K^+$ process, taking into account
the rescatterings in the final $D^-K^+$ and $D^+D^-$ channels, as well as
additional Breit-Wigner resonances representing the $X_1(2900)$, $\psi(3770)$,
and $\chi_{c2}(3930)$ states. The investigation will explore the impact of
rescatterings to the invariant mass spectra and Daltiz plot, and and delve into the nature of the $X_0(2900)$ and $\chi_{c0}(3930)$
resonances. The article will be concluded with a summary of the findings.

\section{Formalism of amplitude for  $B^+$$\to$$D^+ D^- K^+$ process}\label{sec2}

In the current work, we will consider two rescatterings for the
three-body decay process $B^+$$\to$$D^+ D^- K^+$: $D^{*-}K^{*+}$$-$${D^-}{K^+}$ and
${D_s}^+{D_s}^-$$-$${D^+}{D^-}$. These two two-body rescatterings can be described
by the one-boson-exchange model, and the resulting scattering amplitudes will be
incorporated into the three-body decay to obtain the total decay amplitude.

\subsection{Mechanism for $D^{*-}K^{*+}$$-$${D^-}{K^+}$ rescattering}  \label{sec2a}

\begin{figure}[h!]
\includegraphics[bb=100 470 480 660, clip,scale=0.5]{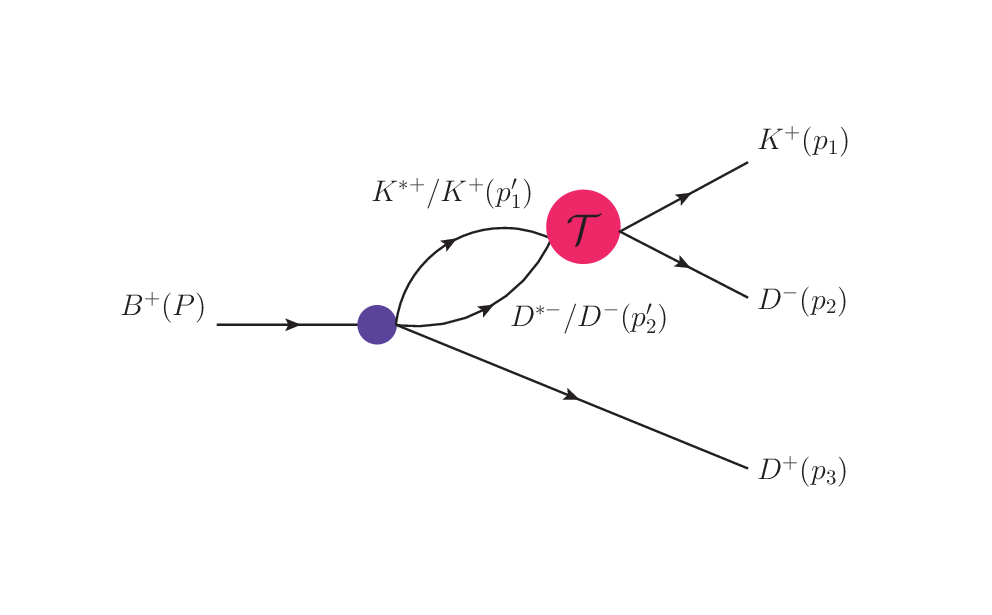}
\caption{ The diagram for process $B^+$$\to$$D^+ D^- K^+$ with $D^{*-}K^{*+}-{D^-}{K^+}$ rescattering. \label{Fig: diagram2900}}
\end{figure}

The experimental data analysis suggests that a resonance structure $X_0(2900)$
can be found in the $D^- K^+$ invariant mass spectrum of the process $B^+$$\to$$
D^+ D^- K^+$. In our model, the $X_0(2900)$ state is interpreted as an S-wave
$D^{*-}K^{*+}$ molecular state. The $B$ meson is assumed to decay first into
$D^{(*)-}$, $K^{(*)+}$, and $D^+$, as depicted in the blue circle in Fig.\ref{Fig:
diagram2900}. The intermediate $D^{(*)-}K^{(*)+}$ state then undergoes a
rescattering process, producing the final $D^-$ and $K^+$ particles, as shown in
the red circle in Fig.\ref{Fig: diagram2900}. When calculating the rescattering
amplitude $\mathcal{T}$, the $D^{*-}K^{*+}$ interaction and its coupling to $D^-
K^+$ are considered.

First, we need to deal with the direct vertex $B^+$ $\to$ $D^+ D^{(*)-}
K^{(*)+}$, as shown in the blue full circle in Fig.~\ref{Fig: diagram2900}.
Following Ref.~\cite{Braaten:2004fk}, the amplitude of the three-body decay can
be constrained by Lorentz invariance. For  direct decay $B^+$$\to$$D^+ D^- K^+$,
the amplitude has the form $A_{B^+ \to D^+ D^- K^+} = c_1$, and for direct decay
$B^+ \to D^+ D^{*-} K^{*+}$, it takes the form $A_{B^+ \to D^+ D^{*-} K^{*+}} =
c_2 \epsilon_{D^{*-}} \epsilon_{K^{*+}}$, where $\epsilon_{D^{*-}}$ and
$\epsilon_{K^{*+}}$ are the polarization vectors of $D^{*-}$ and $K^{*+}$,
respectively. The constants $c_1$ and $c_2$ represent the coupling constants for
the respective channels.  The value of $c_1$ can be obtained by multiplying the
branching fraction of the $B^+$$\to$$D^+ D^- K^+$ with the decay width of the
$B^+$ meson. The branching fraction for $B^+$$\to$$D^+ D^- K^+$ is
$\mathcal{B}_{B^+ \to D^+ D^- K^+} = (2.2 \pm 0.7) \times
10^{-4}$~\cite{LHCb:2022dvn}. However, the branching fraction for $B^+$$\to$$D^+
D^{-} K^{+}$ is not yet available. Therefore, we use $c_1 = 5.397 \times
10^{-5}$, as mentioned in Ref.~\cite{Ding:2023yuo}, and treat $c_2$ as a free
parameter to estimate the decay process due to the lack of experimental
information.  The determination of $c_2$  will be refined based on the
experimental data for the process $B^+$ $\to$ $D^+ D^{(*)-} K^{(*)+}$.

The next step is to construct the potential kernel ${\cal T}$ for the
rescattering process, as shown in Fig.~\ref{Fig: diagram2900}, to find the pole
in the complex energy plane within the
qBSE approach and to calculate the invariant mass spectrum. The one-boson
exchange model will be adopted, with light mesons $\pi$, $\eta$, $\eta'$,
$\rho$, and $\omega$ mediating the interaction between $D^{(*)-}$ and $K^{(*)+}$
mesons. For the systems considered in this
work, the couplings of exchanged light mesons to charmed and strange
mesons are required. Thus, the hidden-gauge Lagrangians with SU(4) symmetry are
suitable for constructing the potential, which reads~\cite{Bando:1984ej,Bando:1987br,Nagahiro:2008cv}
\begin{align} \label{Eq: lagrangianDK}
 \mathcal{L}_{\mathcal{PP}\mathcal{V}} &=-ig~ \langle V_\mu[\mathcal{P},\partial^\mu \mathcal{P}]\rangle,\nonumber\\
 \mathcal{L}_{\mathcal{VV}\mathcal{P}} &=\frac{G'}{\sqrt{2}}~\epsilon^{\mu\nu\alpha\beta}\langle\partial_\mu \mathcal{V}_\nu \partial_\alpha \mathcal{V}_\beta \mathcal{P}\rangle, \nonumber\\
 \mathcal{L}_{\mathcal{VV}\mathcal{V}}&=ig ~\langle (\mathcal{V}_\mu\partial^\nu \mathcal{V}^\mu-\partial^\nu \mathcal{V}_\mu \mathcal{V}^\mu) \mathcal{V}_\nu\rangle,
\end{align}
with $G'={3g'^2}/{4\pi^2f_{\pi}}$, $g'=-{G_{\mathcal{V}}m_{\rho}}/{\sqrt{2}{f_{\pi}}^2}$, $G_\mathcal{V}\simeq 55$ MeV and $f_\pi=93$ MeV and the coupling constant $g=M_\mathcal{V}/{2f_{\pi}}$, $M_\mathcal{V}\simeq 800$ MeV~\cite{Nagahiro:2008cv}.
The $\mathcal{P}$ and $\mathcal{V}$ are the pseudoscalar and vector matrices under SU(4) symmetry as
\begin{equation}
{{\mathcal{P}}} =
\left(
\begin{array}{cccc}
 \frac{\sqrt{3}\pi^0+\sqrt{2}\eta+\eta'}{\sqrt{6}} & \pi^+ & K^+ & \bar{D}^0\\
\pi^- &  \frac{-\sqrt{3}\pi^0+\sqrt{2}\eta+\eta'}{\sqrt{6}}  & K^0 & D^- \\
K^- & \bar{K}^0 & \frac{-\eta+\sqrt{2}\eta'}{\sqrt{3}} & D_s^- \\
D^0 & D^+ & D_s^+ & \eta_c\\
\end{array}
\right),\, \label{PmatrixDK}
\end{equation}
and
\begin{equation}
{\mathcal{V}} =
\left(
\begin{array}{cccc}
 \frac{\rho^0+\omega}{\sqrt{2}} & \rho^+ & K^{* +} & \bar{D}^{* 0} \\
\rho^- & \frac{-\rho^0+\omega}{\sqrt{2}}
 & K^{* 0} & {D}^{* -} \\
K^{* -} & \bar{K}^{* 0} & \phi & D_s^{* -} \\
D^{* 0} & D^{* +} & D_s^{* +} & J/\psi\\
\end{array}
\right)\, .\label{VmatrixDK}
\end{equation}

 \subsection{Mechanism for $D_s^+D_s^--{D^+}{D^-}$ rescattering} \label{sec2b}

\begin{figure}[h!]\begin{center}
\includegraphics[bb=125 450 480 660,clip,scale=0.5]{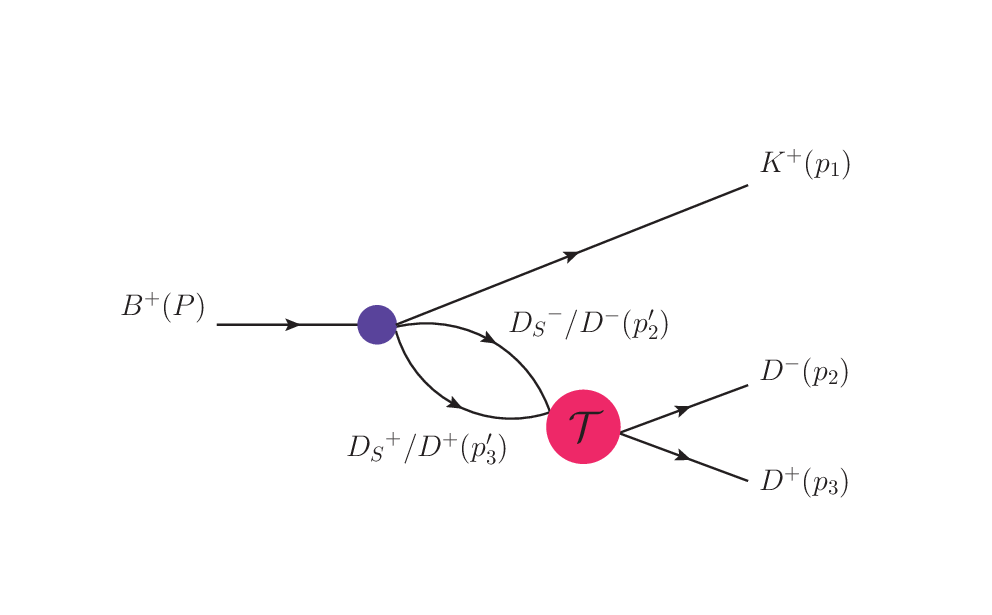}
\end{center}\caption{The diagram for process $B^+\to D^+ D^- K^+$ with $D_s^+D_s^--{D^+}{D^-}$ rescattering.\label{Fig: diagram3930}}
\end{figure}

In this work, we explore the $\chi_{c0}(3930)$ resonance structure in the
${D^+}{D^-}$ invariant mass spectrum, which can be related to
$D_s^+D_s^--{D^+}{D^-}$ rescattering~\cite{Ding:2023yuo}. The Feynman
diagram of such processes is illustrated in Fig.~\ref{Fig: diagram3930}. The
$B^+$ meson decays to $D_{(s)}^+D_{(s)}^-$ and $K^+$ first, and the
intermediate $D_{(s)}^+D_{(s)}^-$ channel will be involved in the
rescattering process, subsequently obtaining the final product $D^+D^-$.  The
amplitude of direct $B^+$$\to$$D_{(s)}^+ D_{(s)}^- K^+$, as shown as the blue full
circle in Fig.~\ref{Fig: diagram3930}, can be written as ${\cal M}_{B^+\to
D_{(s)}^+ D_{(s)}^- K^+} = c_3(c_1)$, where $c_3= 6.027 \times 10^{-5}$ and
$c_1 = 5.397 \times 10^{-5}$ as mentioned in Ref.~\cite{Ding:2023yuo} and
Sec.~\ref{sec2a}.

For the S-wave isoscalar $D_s^+D_s^-$ and ${D^+}{D^-}$ states, the wave functions can be constructed as
\begin{align}
|X_{D\bar{D}}^0\rangle&=\frac{1}{\sqrt{2}}\left(|D^{+}\bar{D}^{-}\rangle+|D^0\bar{D}^{0}\rangle\right), \ 
|X_{D_{s}\bar{D}_{s}}^0\rangle=|D_{s}^{-}D_{s}^{+}\rangle, 
\end{align}

The one-boson-exchange model is used to obtain the interaction between the heavy
mesons. Contributions from vector mesons ($\mathbb{V}=\rho$ and $\omega$ for
${D^+}{D^-}-{D^+}{D^-}$ interaction, $\mathbb{V}=\phi$ for
${D_s}^+{D_s}^--{D_s}^+{D_s}^-$ interaction, and $\mathbb{V}=K^*$ for
${D^+}{D^-}-{D_s}^+{D_s}^-$ interaction) and scalar meson ($\sigma$) exchange
are considered.
Different from the $D^{(*)-}K^{(*)+}$ channel, two charmed mesons are involved here.
Hence, heavy quark symmetry is more suitable to describe the interaction. The
heavy quark effective Lagrangian for heavy mesons interacting with light mesons
reads~\cite{Cheng:1992xi,Yan:1992gz,Wise:1992hn,Burdman:1992gh,Casalbuoni:1996pg},
\begin{align}
  \mathcal{L}_{\mathcal{PP}\mathbb{V}}
  &= -\sqrt{2}\beta{}g_V\mathcal{P}^{}_b\mathcal{P}_a^{\dag}
  v\cdot{\mathbb V}_{ba}
 +\sqrt{2}\beta{}g_V\widetilde{\mathcal{P}}^{\dag}_a
  \widetilde{\mathcal{P}}^{}_b
  v\cdot{\mathbb V}_{ab},\nonumber\\
  \mathcal{L}_{\mathcal{PP}\sigma}
  &= -2g_s\mathcal{P}^{}_b\mathcal{P}^{\dag}_b\sigma
 -2g_s\widetilde{\mathcal{P}}^{}_b\widetilde{\mathcal{P}}^{\dag}_b\sigma,\label{Eq:lagrangianDD}
\end{align} 
where the velocity $v$ should be replaced by
$i\overleftrightarrow{\partial}/2\sqrt{m_i m_f}$ with $m_{i,f}$ being the mass
of the initial or final heavy meson. ${\mathcal{P}}^{(*)T} = (D^{(*)0}, D^{(*)+},
D_s^{(*)+})$. The $\mathbb{V}$ denotes the SU(3) vector matrix, whose elements
are the first $3\times3$ elements of the SU(4) vector matrix in Eq.~(\ref{VmatrixDK}). The parameters
involved here were determined in the literature as $\beta=0.9$, $g_s=0.76$, and
$g_V = 5.9$~\cite{Falk:1992cx,Isola:2003fh,Liu:2009qhy,Chen:2019asm}.

Besides, contribution from the $J/\psi$ exchange is also considered in the
current work since it is found important in the interaction between charmed and anticharmed
mesons~\cite{Ding:2023yuo}. The Lagrangians are written with the help of heavy quark effective
theory as \cite{Casalbuoni:1996pg,Oh:2000qr}, 
\begin{align} {\cal L}_{D_{(s)}
\bar{D}_{(s)}J/\psi} &= ig_{D_{(s)}D_{(s)}\psi} \psi \cdot
\bar{D}\overleftrightarrow{\partial}D,\label{Eq:jpsi} \end{align} where the
couplings are related to a single parameter $g_2$ as
${g_{D_{(s)}D_{(s)}\psi}}/{m_D}= 2 g_2 \sqrt{m_\psi }$,with
$g_2={\sqrt{m_\psi}}/({2m_Df_\psi})$ and $f_\psi=405$ MeV. 

\subsection{Potential kernel} \label{sec2c}

With the above Lagrangians, the potential kernel  in the one-boson-exchange model can be constructed using the
standard Feynman rules, expressed as
\begin{align}
&{\cal V}_{P,\sigma}=I_{{i}}\Gamma_1\Gamma_2 P_{{P},\sigma}f_{P,\sigma}^2(q^2),\nonumber\\
&{\cal V}_{V}=I_{{i}}\Gamma_{1\mu}\Gamma_{2\nu}  P^{\mu\nu}_{V}f_{V}^2(q^2),
\end{align}
for pseudoscalar ($P$), scalar ($\sigma$), and vector ($V$) exchange,
respectively.  The $\Gamma_1$ and $\Gamma_2$ are for the upper and lower
vertices of the one-boson-exchange Feynman diagram, respectively.  The  $I_{i}$
is the flavor factor for certain meson exchange which can be derived using the
Lagrangians in Eqs.~(\ref{Eq: lagrangianDK}), (\ref{Eq:lagrangianDD}) and
(\ref{Eq:jpsi}) and the matrices in Eqs.~(\ref{PmatrixDK}) and (\ref{VmatrixDK})~\cite{He:2015mja}.  The explicit values are
$I_{\pi}=-3/2$, $I_{\eta}=0$, $I_{\eta'}=1/2$, $I_{\rho}=-3/2$ and
$I_{\omega}=1/2$ for $D^{*-}K^{*+}$$-$${D^-}{K^+}$ rescattering, and
$I_{\rho}=3/2$,  $I_{\omega}=1/2$, $I_{\sigma}=I_{J/\psi}=I_{\phi}=1$ for
 $D_s^+D_s^--{D^+}{D^-}$ rescattering, respectively.  The propagators
are defined as usual as
\begin{align}
P_{{P,\sigma}}= \frac{i}{q^2-m_{{P},{\sigma}}^2},\ \
P^{\mu\nu}_{V}=i\frac{-g^{\mu\nu}+q^\mu q^\nu/m^2_{V}}{q^2-m_{V}^2},
\end{align}
where  $q$ is the momentum of  exchanged meson and $m_{V,P,\sigma}$ represents
the mass of the exchanged meson.  
We introduce a form factor $f(q^2)=\Lambda_e^2/(q^2-\Lambda_e^2)$ with a cutoff 
$\Lambda_e$  to compensate for the off-shell effect of the exchanged meson. This
form factor type was also adopted in Ref.~\cite{Gross:2008ps} for studying
nucleon-nucleon scattering with spectator approximation, similar to the current
work. It helps in avoiding overestimation of the contribution of $J/\psi$
exchange in $D_s^+D_s^--{D^+}{D^-}$ rescattering in the current work.

 \subsection{Rescattering amplitudes in qBSE approach} \label{sec2d}

The amplitude of the rescattering process will be expressed within the qBSE
approach. Note that the diagrams illustrated in Sec.\ref{sec2a} and
Sec.\ref{sec2b} involve the same initial and final particles but different
intermediate particles. Here, we present only the rescattering amplitude of the
system mentioned in Sec.\ref{sec2a}, where the final particles $K^+$, $D^-$, and
$D^+$ are labeled as particles 1, 2, and 3, respectively. The rescattering
amplitude for the system discussed in Sec.\ref{sec2b} can be obtained
analogously.

Based on the potential of the interactions constructed above, the rescattering
amplitude can be obtained using the qBSE
approach~\cite{He:2014nya,He:2015mja,He:2017lhy,He:2015yva,He:2017aps}. After
the partial-wave decomposition, the qBSE can be reduced to a 1-dimensional
equation for scattering amplitude ${{\cal T}}^{J^P}$ with a spin-parity $J^P$ as~\cite{He:2015mja},
\begin{align}
&i{{\cal T}}^{J^P}_{\lambda'_1,\lambda'_2,\lambda_1,\lambda_2}({\rm p}',{\rm p})\nonumber\\
&=i{{\cal V}}^{J^P}_{\lambda'_1,\lambda'_2,\lambda_1,\lambda_2}({\rm p}',{\rm
p})+\frac{1}{2}\sum_{\lambda''_1,\lambda''_2}\int\frac{{\rm
p}''^2d{\rm p}''}{(2\pi)^3}\nonumber\\
&\cdot
i{{\cal V}}^{J^P}_{\lambda'_1,\lambda'_2,\lambda''_1,\lambda''_2}({\rm p}',{\rm p}'')
G_0({\rm p}'')i{{\cal T}}^{J^P}_{\lambda''_1,\lambda''_2,\lambda_1,\lambda_2}({\rm p}'',{\rm
p}),\quad\quad \label{Eq: BS_PWA}
\end{align}
where the indices $\lambda'_1$, $\lambda'_2$, $\lambda''_1$, $\lambda''_2$, $\lambda_1$, $\lambda_2$ represent the
helicities of the two rescattering constituents for the final, intermediate, and
initial particles 1 and 2, respectively. $G_0({\bm p}'')$ is a reduced propagator written
in the center-of-mass frame, with $P=(M,{\bm 0})$ as
\begin{align}
	G_0&=\frac{\delta^+(p''^{~2}_2-m_2^{2})}{p''^{~2}_1-m_1^{2}}\nonumber\\
  &
  =\frac{\delta^+(p''^{0}_2-E_2({\bm p}''))}{2E_2({\bm p''})[(W-E_2({\bm
p}''))^2-E_1^{2}({\bm p}'')]},
\end{align}
where $m_{1,2}$ is the mass of particle 1 or 2. As required by the spectator approximation, the heavier particle (particle 2 here)  
is put on shell, with a four-momentum of $p''^0_2$$=$$E_{2}({\rm p}'')$$=$$\sqrt{
m_{2}^{~2}+\rm p''^2}$. The corresponding four-momentum for the lighter particle
(particle 1 here) $p''^0_1$ is then $W-E_{2}({\rm p}'')$ with $W$ being the
center-of-mass energy of the system. Here and hereafter we define the value of
the momentum ${\rm p}=|{\bm p}|$. And the momentum of particle 1  ${\bm
p}''_1=-{\bm p}''$ and the momentum of particle 2 ${\bm p}''_2={\bm p}''$. 

The partial wave potential ${\cal
V}_{\lambda'_1,\lambda'_2,\lambda_1,\lambda_2}^{J^P}$ can be obtained from the
potential as
\begin{align}
i{\cal V}_{\lambda'_1,\lambda'_2,\lambda_1,\lambda_2}^{J^P}({\rm p}',{\rm p})
&=2\pi\int d\cos\theta
~[d^{J}_{\lambda_{21}\lambda'_{21}}(\theta)
i{\cal V}_{\lambda'_1\lambda'_2,\lambda_1\lambda_2}({\bm p}',{\bm p})\nonumber\\
&+\eta d^{J}_{-\lambda_{21}\lambda'_{21}}(\theta)
i{\cal V}_{\lambda'_1,\lambda'_2,-\lambda_1,-\lambda_2}({\bm p}',{\bm p})],\label{Eq:PWAV}
\end{align}
where $\eta=PP_1P_2(-1)^{J-J_1-J_2}$ with $P$ and $J$ being parity and spin for
system and constituent 1 or 2. $ \lambda_{21}=\lambda_2-\lambda_1$. The initial
and final relative momenta are chosen as ${\bm p}'=(0,0,{\rm p}')$  and ${\bm
p}=({\rm p}\sin\theta,0,{\rm p}\cos\theta)$. The $d^J_{\lambda'\lambda}(\theta)$
is the Wigner d-matrix. An  exponential regularization  is also introduced as
a form factor into the reduced propagator as $G_0({\rm p}'')\to G_0({\rm
p}'')e^{-2(p''^2_2-m_2^2)^2/\Lambda_r^4}$~\cite{He:2015mja}. The cutoff
parameter $\Lambda_r$ and $\Lambda_e$ can be chosen as different values, 
but have a similar effect on the result. For simplification, we set $\Lambda_r=\Lambda_e$ in the current work.

The amplitude ${\cal T}$ can be determined by discretizing the momenta ${\rm
p}'$, ${\rm p}$, and ${\rm p}''$  in the integral equation~(\ref{Eq: BS_PWA}) 
using Gauss quadrature with a weight  $w({\rm p}_i)$.  After this discretization, 
the integral equation can be reformulated as a matrix equation  ~\cite{He:2015mja}
\begin{align}
{T}_{ik}
&={V}_{ik}+\sum_{j=0}^N{ V}_{ij}G_j{T}_{jk}.\label{Eq: matrix}
\end{align}
Here, the propagator $G$ is represented as a diagonal matrix:
\begin{align}
	G_{j>0}&=\frac{w({\rm p}''_j){\rm p}''^2_j}{(2\pi)^3}G_0({\rm
	p}''_j), \nonumber\\
G_{j=0}&=-\frac{i{\rm p}''_o}{32\pi^2 W}+\sum_j
\left[\frac{w({\rm p}_j)}{(2\pi)^3}\frac{ {\rm p}''^2_o}
{2W{({\rm p}''^2_j-{\rm p}''^2_o)}}\right],
\end{align}
where the on-shell momentum is given by
\begin{align}{\rm p}''_o=\frac{1}{2W}\sqrt{[W^2-(m_1+m_2)^2][W^2-(m_1-m_2)^2]}.\label{Eq: mometum onshell}
\end{align}
To identify the pole of the rescattering amplitude  in the energy complex plane, we seek the position where 
$|1-{ V}G|=0$ with $z=E_R + i\Gamma/2$ corresponding to the total energy and width.

After incorporating the amplitudes of the direct decay and rescattering processes, 
the total amplitude of the process $B^+$$\to$$D^+ D^- K^+$ with $D^{*-}K^{*+}$$-$${D^-}{K^+}$ 
rascattering  can be expressed in the center-of-mass frame of particles 1 and 2 as follows~\cite{He:2017lhy,Ding:2023yuo}:
\begin{align}
{\cal M}(p_1,p_2,p_3)
&=\sum_{\lambda'_1,\lambda'_2}\int \frac{d^4p'^{cm}_2}{(2\pi)^4} 
{\cal T}_{\lambda'_1,\lambda'_2}(p^{cm}_1,p^{cm}_2;p'^{cm}_1,p'^{cm}_2) \nonumber\\
&\cdot \ G_0(p'^{cm}_2){\cal A}_{\lambda'_1,\lambda'_2}(p'^{cm}_1,p'^{cm}_2,p_3^{cm}).
\end{align}
Here, the helicities of the initial and final particles of the process 
$B^+\to D^+ D^- K^+$ have been omitted since they are all zero. 
The momenta with the superscript $cm$ refer to those in the center-of-mass frame of particles 1 and 2. 

To analyze the direct decay amplitude
${\cal A}_{\lambda'_1,\lambda'_2}$ a partial wave expansion is required, similar to the expansion performed on the rescattering potential kernel
${\cal V}$ conducted in Eq.~(\ref{Eq:PWAV}), as shown in Ref.~\cite{Gross:2008ps},
\begin{align}
{\cal A}_{\lambda'_1,\lambda'_2}(p'^{cm}_1,p'^{cm}_2,p_3^{cm})
=\sum_{J\lambda'_{1}\lambda'_{2}}N_JD^{J*}_{\lambda'_{1},\lambda'_{2}}( \Omega^{cm}_2) 
{\cal A}^{J}_{\lambda'_1,{\lambda}^{'}_{2}}({\rm p}'^{cm}_2,p_3^{cm}),\label{Eq: decay}
\end{align}
where $N_J$ is a normalization constant with the value of $\sqrt{(2J+1)/4\pi}$,
and $\Omega_2^{cm}$ is the spherical angle of momentum of particle 2. 
Hence, the partial-wave amplitude for $J^P=0^+$ is given by
\begin{align}
{\cal M}^{0^+}(p_1,p_2,p_3)&=\frac{1}{2}N_{0}\sum_{\lambda'_{1}\lambda'_{2}}
\int \frac{{\rm p}'^{cm2}_2d{\rm p}'^{cm}_2}{(2\pi)^3}
i{\cal T}^{0^+}_{\lambda'_1,\lambda'_2}({\rm p}'^{cm}_2,{\rm p}^{cm}_2)
\nonumber\\ 
&\ \ \cdot\ \  G_0({\rm p}'^{cm}_2) 
{\cal A}^{0^+}_{\lambda'_1,\lambda'_2}({\rm p}'^{cm}_2,p_3^{cm}).
\end{align}

\section{The numerical results}\label{sec3}

In this section, we will calculate the $D^-K^+$ and $D^-D^+$ invariant mass
spectra and generate the corresponding Dalitz plot  using Monte Carlo simulation.
Our analysis aims to compare these results with experimental data to investigate the nature of the $X_0(2900)$ and $\chi_{c0}(3930)$ states.

\subsection{ Invariant mass spectrum and Dalitz plot} \label{sec2e}

With the preparation in the previous section, we are able to calculate the decay amplitude of the process  
$B^+$$\to$$D^+ D^- K^+$  with rescattering of $D^{*-}K^{*+}$$-$$D^{-}K^{+}$ with $J^P=0^+$
for the $X_0(2900)$ resonance.  Similarly, we can obtain the decay amplitude for 
$D_{s}^+D_{s}^--{D}^+{D}^-$ with $J^P=0^+$ for $\chi_{c0}(3930)$ resonance using a similar approach.

In the current works, we focus on the roles played by rescatterings and relevant
molecular states process $B^+\to D^+ D^- K^+$. However, it is important to note that
if we only consider rescatterings related to the $X_0(2900)$ and $\chi_{c0}(3930)$
resonances, the model amplitude may be too simplistic to accurately replicate the 
invariant mass spectra observed in experiments. 
To address this, we introduce Breit-Wigner resonances near 
2900~MeV with $J=1$, near 3770~MeV with $J=1$, and  near 3930~MeV
with $J=2$ to account for the $X_1(2900)$, $\psi(3770)$, and $\chi_{c2}(3930)$ signals observed by
the LHCb Collaboration. The amplitude model can be expressed as,
\begin{align}
A(J)=a_{J}B W(M_{ab}) \times T(\Omega).\label{Eq: BW}
\end{align}
Here $a_{J}$ is a free parameter.  The relativistic Breit-Wigner function 
$BW(M_{ab})$ is defined as,
\begin{align}
B W\left(M_{ab}\right)=\frac{F_r F_D}{M_r^2-M_{ab}^2-i \Gamma_{ab} M_r},
\end{align}
where $F_r$ and $F_D$ represent the Blatt-Weisskopf damping factors for the $B$ meson
and the resonance, respectively. $M_r$ is the mass of resonance, $M_{ab}$ is the invariant
mass while $ab$ denotes $12$ or $23$ depending on the system we choose, and
$\Gamma_{ab}$ is the mass-dependent width, which can be expressed as
\begin{align}
\Gamma_{ab}&=\Gamma_r\left(\frac{p_{ab}}{p_r}\right)^{2 J+1}\left(\frac{M_r}{M_{ab}}\right) F_r^2.\nonumber\\
p_{ab}&=\frac{\sqrt{\left(M_{ab}^2-m_a^2-m_b^2\right)^2-4 m_a^2 m_b^2}}{2 M_{ab}} ,\label{Eq: Gammaab}
\end{align}
where $\Gamma_r$ and $J$ are the width and the spin of the resonance. The
quantity $p_{ab}$ is the momentum of either daughter in the $ab$ rest frame, and
$p_r$ is the value of $p_{ab}$ when $M_{ab}=M_r$. The exact expressions of the
Blatt-Weisskopf factors~\cite{Blatt:1952ije} are given in
Ref.~\cite{BaBar:2010wqe}. The angular term $ T(\Omega)$ is also given in
Ref.~\cite{BaBar:2010wqe} and depends on the masses of the particles involved in
the reaction as well as on the spin of the intermediate resonance.

The parameters associated with the additional $X_1(2900)$, $\psi(3770)$ and
$\chi_{c2}(3930)$ resonances are predetermined based on the 
experimental values~\cite{LHCb:2020bls,LHCb:2020pxc}. The predetermined masses and
widths are provided in Table~\ref{LHCbdata}.  Additionally, the resonances
resulting from the rescatterings correspond to poles in the complex energy
plane. By adjusting the masses and widths to better match the invariant mass
spectra, the values for the masses and widths of $X_0(2900)$ and
$\chi_{c0}(3930)$ are also determined and included in Table~\ref{Tpara}. 
Further discussion on these values will be provided later.
\renewcommand\tabcolsep{0.25cm}
\renewcommand{\arraystretch}{1.2}
\begin{table}[h!]
\centering
\caption{Masses and widths for the resonances involved. 
 Predetermined values are cited from Refs.~\cite{LHCb:2020bls,LHCb:2020pxc}, and the 
 fitted values are determined by fitting the invariant mass spectra.
\label{Tpara}}
%%%%%%%%%%%%
\begin{tabular}{l ccc }
\hline
&Resonance & {Mass (MeV)} & Width (MeV) \\
\hline
Predetermiend 
&$X_1(2900)$ & 2904.0  & 110.0\\
&$\psi(3770)$ & 3778.1  &\ \ \ \ 0.9\\
&$\chi_{c2}(3930)$ & 3926.8 & \ \ 34.2 \\\hline
Fitted &$X_0(2900)$ & 2884.9 &\ \ 62.0 \\
&$\chi_{c0}(3930)$ & 3923.9  & \ \ \ \ 0.1 \\
\hline
\end{tabular}
%%%%%%%%%%%%
\end{table}

Besides, a parameterized background contribution is introduced into
the $D^-K^+$ invariant mass distribution from rescattering described 
in Sec.~\ref{sec3a}, which can be written as
\begin{align}
\hat{\cal A}^{bk}(M_{12})=e^{d(i\pi)}c(M_{12}-M_{min})^a(M_{max}-M_{12})^b,\label{Eq: background}
\end{align}
where the parameters for the background are chosen as $(a, b, c, d)=(4.0,0.5,5.6,0.5)$ to fit the experimental data.

The total decay width, incorporating all these contributions to the amplitude ${\cal M}(p_1, p_2, p_3)$, can be expressed as
\begin{align}
{d\Gamma}&=\frac{(2\pi)^4}{2M_{B}}|{\cal M}(p_1,p_2,p_3)|^2d\Phi_3,\label{Eq: IM} 
\end{align}
where $M_B$ are the mass of initial $B^+$ meson. In this work, the phase space $d\Phi_3$ in Eq.~(\ref{Eq: IM})  is obtained using the 
GENEV code in FAWL, which employs the Monte Carlo method to generate events of the 
three-body final state. The phase space is defined as, 
\begin{align}
R_3=(2 \pi)^5 d \Phi_3=\prod_i^3 \frac{d^3 p_i}{2 E_i} \delta^4\left(\sum_i^n p_i-P\right),
\end{align}
where $p_i$ and $E_i$ representing  the momentum and energy of the final particle 
$i$ are generated by the code . By simulating  $5\times10^5$ events, the event distribution can be obtained, allowing for the 
visualization of the Dalitz plot and the invariant mass spectra with respect to $m_{D^-K^+}$ and
$m_{D^+D^-}$.

It is important to note that an overall scaling factor for the invariant mass
spectra cannot be determined due to the absence of information on the total
number of $B^+$ candidates, which was not provided by the LHCb collaboration.
Therefore, it is necessary to scale the theoretical decay distribution to the
experimental data to facilitate a comparison between theoretical predictions and
experimental results. It is crucial to emphasize that this scaling factor
renders only the relative values of the coupling constants $c_1$, $c_2$ and
$c_3$ meaningful.  Additionally, the cutoff parameter $\Lambda_e$  is fine-tuned
to best match the experimental data.  Given that the interaction mechanisms
differ between the $D^{*-}K^{*+}-{D^-}{K^+}$ rescattering and
${D_s}^+{D_s}^--{D^+}{D^-}$ rescattering, it is reasonable to set different cutoff values of $\Lambda_e$ for the two rescatterings, 3.3 and 1.8 GeV, respectively, to ensure a
good fit to the experimental data. 

\subsection{$D^-K^+$ invariant mass spectrum and $X_{0,1}(2900)$} \label{sec3a}

\begin{figure}[h!]
  \includegraphics[bb=65 85 440 278,clip,scale=0.78]{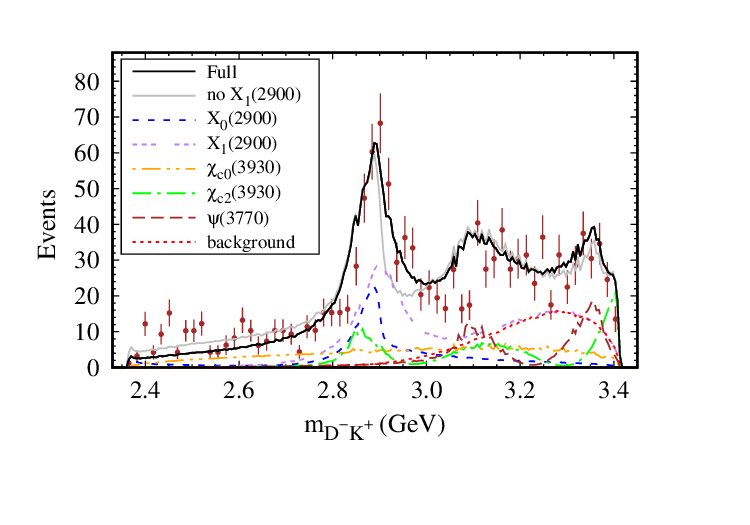}
    \caption{The $D^-K^+$ invariant mass spectrum for process $B^+\to D^+ D^- K^+$. 
    The black (grey) full curve shows the contribution from the total
  amplitude (without $X_1(2900)$), and the blue, orange, green, brown, purple,
  and red dashed curves show the contributions from $X_{0}(2900)$,
  $\chi_{c0}(3930)$, $\chi_{c2}(3930)$, $\psi(3770)$, $X_{1}(2900)$, and
  background, respectively.  The red points with error bars are the data from
  the LHCb experiment~\cite{LHCb:2020pxc}. The results are obtained with $5\times10^5$ simulations.\label{Fig: ims12}}
\end{figure}

As shown in Fig.~\ref{Fig: ims12}, the LHCb data suggest an obvious resonance
structure around 2900 MeV in the $D^-K^+$ invariant mass spectrum of the process
$B^+$$\to$$D^+ D^- K^+$. Such a structure is close to the $D^*K^*$ threshold and
was explained as $D^{*-}K^{*+}$ molecular states in the literature. As we can
see from the blue dashed curve representing the contribution from
$D^{*-}K^{*+}$$-$${D^-}{K^+}$ rescattering in Fig.~\ref{Fig: ims12}, an obvious peak
is produced near 2885 MeV as expected, which can be related to the $X_0(2900)$
resonance. However, a satisfactory fit of the LHCb's data cannot be obtained if
the $X_1(2900)$ is not included, as shown by the gray line in Fig.~\ref{Fig:
ims12}, because it seems narrow and located to the left compared to the
experimental structure. In the experimental article~\cite{LHCb:2020pxc}, the
structure observed near 2900 MeV was suggested to be formed by two states,
$X_0(2900)$ and $X_1(2900)$, with $X_0(2900)$ having a smaller mass and width.
Thus, we introduce a Breit-Wigner resonance near 2900 MeV with $J=1$ to fit the
$X_1(2900)$ structure, which can be written as Eqs.~(\ref{Eq: BW})-(\ref{Eq:
Gammaab}) with predetermined mass and width listed in Table~\ref{LHCbdata}.

A good description of the region near 2900~MeV is obtained after including the
$X_1(2900)$ contribution, together with reflections from the $D^+D^-$
structures. The peak near 2900~MeV becomes wider and is in better agreement with
the experimental result, as shown by the black curve. Thus, our result still
supports the assumption of $X_0(2900)$ as an S-wave $D^-K^+$ molecular state,
and the $X_1(2900)$ state is necessary to describe the structure near 2900MeV as
well. 

Two bumps can be observed around 3.0 to 3.5~GeV due to the contribution of
$\psi(3770)$ and $\chi_{c2}(3930)$ states, which is consistent with the relevant
data analysis of the LHCb Collaboration~\cite{LHCb:2020pxc}. No clear peak
structure but a background-like signal can be found from the orange dashed curve
representing the $\chi_{c0}(3930)$ state, indicating that this state has no
significant contribution to the $D^{-}K^{+}$ invariant mass spectrum in our
model.

\subsection{${D}^+ {D}^-$ invariant mass spectrum and $\chi_{c0}(3930)$} \label{sec3b}

In the ${D}^+ {D}^-$ invariant mass spectra of the process $B^+\to D^+ D^- K^+$,
two prominent resonance structures are observed around 3770 and 3930 MeV,
respectively. In our previous study, we investigated the $D_{s}^+
D_{s}^-$$-$$D^+D^-$ rescattering within the ${D}^+ {D}^-$ invariant mass
spectrum to explore the origins of $X(3960)$ and $\chi_{c0}(3930)$. Our findings
indicated that the peak attributed to $\chi_{c0}(3930)$ was too narrow to fully
explain the experimental structure around 3930 MeV~\cite{Ding:2023yuo}. Given
that this structure is suggested to arise from two states, $\chi_{c0}(3930)$ and
$\chi_{c2}(3930)$, it is reasonable to propose that $\chi_{c2}(3930)$ plays a
crucial role in forming the resonance near 3930 MeV in the ${D}^+ {D}^-$
invariant mass spectrum. Additionally, a distinct structure
around 3770 MeV is identified as $\psi(3770)$ with spin parity $J^{PC}=1^{--}$
in experimental reports. Therefore, in this study, Breit-Wigner resonances with
$J=2$ near 3930 MeV and $J=1$ near 3770 MeV are introduced to model the
$\chi_{c2}(3930)$ and $\psi(3770)$ structures, respectively. These resonances
are parameterized according to Eqs.~(\ref{Eq: BW})-(\ref{Eq: Gammaab})  with
predetermined mass and width listed in Table~\ref{LHCbdata}.

\begin{figure}[h!]
  \includegraphics[bb=65 85 440 288,clip,scale=0.78]{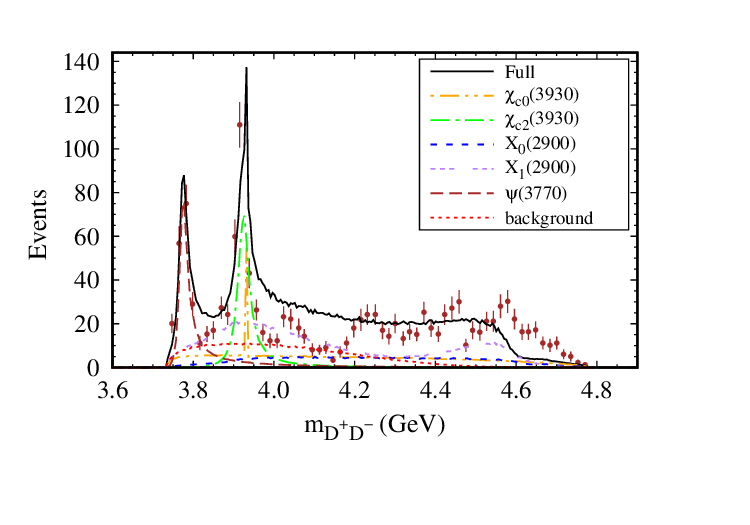}
  \caption{The $D^+D^-$ invariant mass spectrum for process $B^+\to D^+ D^- K^+$. 
  The black solid curve shows the contribution from the total amplitude,
  and the orange, green, brown, blue, purple, and red dashed curves show the
  contributions from $\chi_{c0}(3930)$, $\chi_{c2}(3930)$, $\psi(3770)$,
  $X_0(2900)$, $X_1(2900)$, and the background, respectively. The red points
  with error bars are the data from the LHCb experiment~\cite{LHCb:2020pxc}. The results are obtained with $5\times10^5$ simulations. \label{Fig: ims23}}
  
\end{figure}

The $D^+D^-$ invariant mass spectrum for $B^+\to D^+ D^- K^+$ via the
intermediate states $\chi_{c0}(3930)$, $\chi_{c2}(3930)$, $\psi(3770)$,
$X_0(2900)$, and $X_1(2900)$ can be obtained in Fig.~\ref{Fig: ims23}. As shown
in the orange dashed curve representing the contribution of $\chi_{c0}(3930)$
state in Fig.~\ref{Fig: ims23}, a significant but extremely narrow peak around
3930 MeV can be observed, which can be associated with the rescattering of
$D_{s}^+ D_{s}^-$$-$${D}^+ {D}^-$ channels. After overlapping the Breit-Wigner
resonance introduced to fit the $\chi_{c2}(3930)$ structure, the width of the
peak around 3930 MeV becomes larger as the black curve shows, and the explicit
shape of the experimental structure can be better fitted compared with our
previous work. This result favors the assumption of the $\chi_{c0}(3930)$ as an
S-wave $D^+D^-$ molecular state but may play a minor role in forming the
structure around 3930 MeV in the $D^+D^-$ invariant mass spectrum. 

Additionally, a peak near 3770 MeV is due to the Breit-Wigner resonance introduced to fit the $\psi(3770)$ structure. For structures in the higher energy region, experimental analysis~\cite{LHCb:2020pxc} suggests that additional states, such as $\psi(4040)$, $\psi(4160)$, and $\psi(4415)$, should be included. However, since these states are not the focus of the current work, further fitting of these resonances is not pursued. No significant peak structure is found in the blue dashed curve in Fig.~\ref{Fig: ims23}, indicating that $X_0(2900)$ does not contribute significantly to the $D^+D^-$ invariant mass spectrum.

\subsection{Dalitz Plot} \label{sec3c}

\begin{figure}[h!]
\includegraphics[bb=90 90 360 280,clip,scale=0.92]{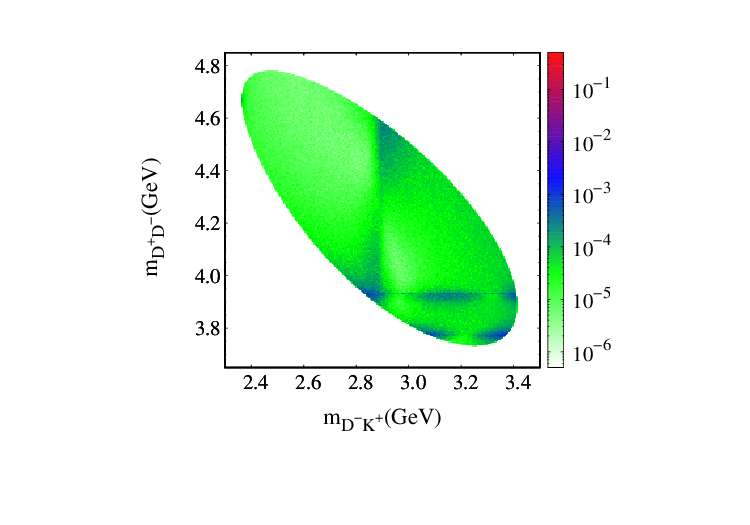}
\caption{The Dalitz plot for the $B^+\to D^+ D^- K^+$ process. The color box represents the ratio of the number of events in a bin of 0.005~GeV $\times$ 0.005~GeV to the total number of events. The results are obtained with $5\times10^5$ simulations.\label{Fig: dp}}
\end{figure}

In the above, the invariant mass spectra  for the process $B^+$$\to$$ D^+ D^- K^+$ is presented. 
To provide a more explicit picture for such process, we present the Dalitz plot
against the invariant masses $m_{D^-K^+}$ and $m_{D^+D^-}$ of the final
particles in the molecular state picture in Fig.~\ref{Fig: dp}.

When comparing our results in Fig.~\ref{Fig: dp} with the LHCb data (Figure 8 in
Ref.~\cite{LHCb:2020pxc}), we observe a striking similarity in the overall
pattern. In the Dalitz plot, two prominent horizontal strips appear at
$m_{D^+D^-}$ around 3.77 and 3.93 GeV, indicating contributions from the
$\psi(3770)$ and $\chi_{c0,2}(3930)$ resonances, respectively. Additionally, a
distinct vertical strip is visible at $m_{D^-K^+}$ around 2900 GeV, which can be
associated with the $X_{0,1}(2900)$ states.
The intermittency of these strips, observed in both our results and the
experimental data from LHCb, provides additional insights. This pattern suggests
that the resonances do not originate solely from scalar states, confirming the
necessity of including the $\chi_{c2}(3930)$ and $X_1(2900)$ states.

\section{Summary}\label{sec4}

In this work, we studied the $D^-K^+$ and $D^+D^-$ invariant mass spectra, as well as the Dalitz plot, for the $B^+ \to D^+ D^- K^+$ process. We focused on the $D^{-}K^{+} - D^-K^+$ and $D_s^+D_s^- - D^+D^-$ rescattering, which can be related to the resonances $X_0(2900)$ and $\chi_{c0}(3930)$ as molecular states with spin parity $0^+$. The theoretical results, calculated using the quasipotential Bethe-Salpeter equation approach, are compared with the experimental data from the LHCb Collaboration to understand the nature of $X_0(2900)$ and $\chi_{c0}(3930)$.

The $D^-K^+$ invariant mass spectrum for the decay process $B^+\to D^+ D^- K^+$
is examined. The $D^{*-}K^{*+}$ interaction produces a significant peak at 2886
MeV with a width of 62 MeV, identified as the $X_0(2900)$ contribution suggested
by LHCb. However, this peak is slightly sharper than the experimental structure.
Since the structure was suggested to be formed by two states, $X_{0}(2900)$ and
$X_{1}(2900)$, the latter is introduced as a Breit-Wigner resonance near 2900
MeV with $J=1$ into our model. Including this state widens the structure,
bringing the calculation into good agreement with the experimental $D^-K^+$
invariant mass spectrum.  To describe the distribution in the higher energy
region, contributions from $\psi(3770)$ and $\chi_{c2}(3930)$ are needed.

The $D^+D^-$ invariant mass spectrum, which includes contributions from the
$\psi(3770)$, $\chi_{c0}(3930)$, and $\chi_{c2}(3930)$ states as suggested by
LHCb, was also investigated. The $\chi_{c0}(3930)$, originating from the
$D_s^+D_s^-$ interaction, exhibits a very narrow peak and plays a smaller role
in forming the experimentally observed peak near 3930 MeV compared to the
$\chi_{c2}(3930)$, which is introduced as a Breit-Wigner resonance.
Additionally, to simulate the $\psi(3770)$ resonance, we introduced a
Breit-Wigner resonance located at 3770 MeV with $J$$=$$1$. The event
distribution in the $D^+D^-$ invariant mass spectrum can be described by two
sharp peaks corresponding to the $\psi(3770)$ and the overlap of
$\chi_{c0}(3930)$ and $\chi_{c2}(3930)$. The distribution at higher energy
regions requires contributions from more states, which is not the focus of the
current work and therefore not considered.

Additionally, the Dalitz plot in the $m_{D^-K^+} - m_{D^+D^-}$ plane for the
process $B^+\to D^+ D^- K^+$ was presented. Obvious intermittent strips are
observed at about 2.90 GeV at $m_{D^-K^+}$ and 3.77 and 3.93 GeV at $m_{D^+D^-}$
in the Dalitz plot. Our calculations support the molecular interpretations for
$X_0(2900)$ and $\chi_{c0}(3930)$ states. However, it was found necessary to
include both spin-0 and spin-1 states in the $X_J(2900)$ region and both spin-0
and spin-2 states in the $\chi_{cJ}(3930)$ region.

\vskip 10pt

\noindent {\bf Data Availability Statement} This manuscript has no associated data or the data will not be deposited. [Authors' comment: This is a theoretical study and no external data are associated with this work.]

%

%\bibliography{../../../reference/Jabref/bibliography}

\begin{thebibliography}{23}%

%\cite{LHCb:2020bls}
\bibitem{LHCb:2020bls}
R.~Aaij \textit{et al.} [LHCb],
``A model-independent study of resonant structure in $B^+\to D^+D^-K^+$ decays,''
Phys. Rev. Lett. \textbf{125} (2020), 242001
%doi:10.1103/PhysRevLett.125.242001
%[arXiv:2009.00025 [hep-ex]].
%189 citations counted in INSPIRE as of 24 Jun 2024

%\cite{LHCb:2020pxc}
\bibitem{LHCb:2020pxc}
R.~Aaij \textit{et al.} [LHCb],
``Amplitude analysis of the $B^+\to D^+D^-K^+$ decay,''
Phys. Rev. D \textbf{102} (2020), 112003
%doi:10.1103/PhysRevD.102.112003
%[arXiv:2009.00026 [hep-ex]].
%226 citations counted in INSPIRE as of 24 Jun 2024

%\cite{LHCb:2024vfz}
\bibitem{LHCb:2024vfz}
R.~Aaij \textit{et al.} [LHCb],
``Observation of new charmonium(-like) states in $B^+ \to D^{*\pm} D^{\mp} K^+$ decays,''
arXiv:2406.03156 [hep-ex]
%1 citations counted in INSPIRE as of 04 Jul 2024

%\cite{Zhang:2020oze}
\bibitem{Zhang:2020oze}
J.~R.~Zhang,
``Open-charm tetraquark candidate: Note on $X_0$(2900),''
Phys. Rev. D \textbf{103} (2021) no.5, 054019
%doi:10.1103/PhysRevD.103.054019
%[arXiv:2008.07295 [hep-ph]].
%52 citations counted in INSPIRE as of 26 Jun 2024

%\cite{Wang:2020xyc}
\bibitem{Wang:2020xyc}
Z.~G.~Wang,
``Analysis of the $X_0(2900)$ as the scalar tetraquark state via the QCD sum rules,''
Int. J. Mod. Phys. A \textbf{35} (2020) no.30, 2050187
%doi:10.1142/S0217751X20501870
%[arXiv:2008.07833 [hep-ph]].
%55 citations counted in INSPIRE as of 26 Jun 2024

%\cite{He:2020jna}
\bibitem{He:2020jna}
X.~G.~He, W.~Wang and R.~Zhu,
``Open-charm tetraquark $X_c$ and open-bottom tetraquark $X_b$,''
Eur. Phys. J. C \textbf{80} (2020) no.11, 1026
%doi:10.1140/epjc/s10052-020-08597-1
%[arXiv:2008.07145 [hep-ph]].
%64 citations counted in INSPIRE as of 26 Jun 2024

%\cite{Wang:2020prk}
\bibitem{Wang:2020prk}
G.~J.~Wang, L.~Meng, L.~Y.~Xiao, M.~Oka and S.~L.~Zhu,
``Mass spectrum and strong decays of tetraquark ${\bar{c}}{\bar{s}} qq$ states,''
Eur. Phys. J. C \textbf{81} (2021) no.2, 188
%doi:10.1140/epjc/s10052-021-08978-0
%[arXiv:2010.09395 [hep-ph]].
%38 citations counted in INSPIRE as of 26 Jun 2024

%\cite{Liu:2020orv}
\bibitem{Liu:2020orv}
X.~H.~Liu, M.~J.~Yan, H.~W.~Ke, G.~Li and J.~J.~Xie,
``Triangle singularity as the origin of $X_0(2900)$ and $X_1(2900)$ observed in $B^+\to D^+ D^- K^+$,''
Eur. Phys. J. C \textbf{80} (2020) no.12, 1178
%doi:10.1140/epjc/s10052-020-08762-6
%[arXiv:2008.07190 [hep-ph]].
%62 citations counted in INSPIRE as of 26 Jun 2024

%\cite{Burns:2020epm}
\bibitem{Burns:2020epm}
T.~J.~Burns and E.~S.~Swanson,
``Kinematical cusp and resonance interpretations of the $X(2900)$,''
Phys. Lett. B \textbf{813} (2021), 136057
%doi:10.1016/j.physletb.2020.136057
%[arXiv:2008.12838 [hep-ph]].
%38 citations counted in INSPIRE as of 26 Jun 2024

%\cite{Molina:2020hde}
\bibitem{Molina:2020hde}
R.~Molina and E.~Oset,
``Molecular picture for the $X_0(2866)$ as a $D^* \bar{K}^*$ $J^P=0^+$ state and related $1^+,2^+$ states,''
Phys. Lett. B \textbf{811} (2020), 135870
[erratum: Phys. Lett. B \textbf{837} (2023), 137645]
%doi:10.1016/j.physletb.2020.135870
%[arXiv:2008.11171 [hep-ph]].
%66 citations counted in INSPIRE as of 26 Jun 2024

%\cite{Chen:2020aos}
\bibitem{Chen:2020aos}
H.~X.~Chen, W.~Chen, R.~R.~Dong and N.~Su,
``$X_0$(2900) and $X_1$(2900): Hadronic Molecules or Compact Tetraquarks,''
Chin. Phys. Lett. \textbf{37} (2020) no.10, 101201
%doi:10.1088/0256-307X/37/10/101201
%[arXiv:2008.07516 [hep-ph]].
%75 citations counted in INSPIRE as of 26 Jun 2024

%\cite{Agaev:2020nrc}
\bibitem{Agaev:2020nrc}
S.~S.~Agaev, K.~Azizi and H.~Sundu,
``New scalar resonance X 0(2900) as a molecule: mass and width,''
J. Phys. G \textbf{48} (2021) no.8, 085012
%doi:10.1088/1361-6471/ac0b31
%[arXiv:2008.13027 [hep-ph]].
%48 citations counted in INSPIRE as of 26 Jun 2024

%\cite{Mutuk:2020igv}
\bibitem{Mutuk:2020igv}
H.~Mutuk,
``Monte-Carlo based QCD sum rules analysis of $X_0$(2900) and $X_1$(2900),''
J. Phys. G \textbf{48} (2021) no.5, 055007
%doi:10.1088/1361-6471/abeb7f
%[arXiv:2009.02492 [hep-ph]].
%25 citations counted in INSPIRE as of 26 Jun 2024

%\cite{Liu:2020nil}
\bibitem{Liu:2020nil}
M.~Z.~Liu, J.~J.~Xie and L.~S.~Geng,
``$X_0(2866)$ as a $D^*\bar{K}^*$ molecular state,''
Phys. Rev. D \textbf{102} (2020) no.9, 091502
%doi:10.1103/PhysRevD.102.091502
%[arXiv:2008.07389 [hep-ph]].
%82 citations counted in INSPIRE as of 26 Jun 2024

%\cite{Xiao:2020ltm}
\bibitem{Xiao:2020ltm}
C.~J.~Xiao, D.~Y.~Chen, Y.~B.~Dong and G.~W.~Meng,
``Study of the decays of $S-$wave $\bar D^\ast K^\ast$ hadronic molecules: The scalar $X_0(2900)$ and its spin partners $X_{J(J=1,2)}$,''
Phys. Rev. D \textbf{103} (2021) no.3, 034004
%doi:10.1103/PhysRevD.103.034004
%[arXiv:2009.14538 [hep-ph]].
%43 citations counted in INSPIRE as of 26 Jun 2024

%\cite{He:2020btl}
\bibitem{He:2020btl}
J.~He and D.~Y.~Chen,
``Molecular picture for $X_0(2900)$ and $X_1(2900)$,''
Chin. Phys. C \textbf{45} (2021) no.6, 063102
%doi:10.1088/1674-1137/abeda8
%[arXiv:2008.07782 [hep-ph]].
%47 citations counted in INSPIRE as of 26 Jun 2024

%\cite{Kong:2021ohg}
\bibitem{Kong:2021ohg}
S.~Y.~Kong, J.~T.~Zhu, D.~Song and J.~He,
``Heavy-strange meson molecules and possible candidates $D_{s0}^*(2317)$, $D_{s1}(2460)$, and $X_0(2900)$,''
Phys. Rev. D \textbf{104} (2021) no.9, 094012
%doi:10.1103/PhysRevD.104.094012
%[arXiv:2106.07272 [hep-ph]].
%30 citations counted in INSPIRE as of 26 Jun 2024

%\cite{LHCb:2022aki}
\bibitem{LHCb:2022aki}
R.~Aaij \textit{et al.} [LHCb],
``Observation of a Resonant Structure near the Ds+Ds- Threshold in the $B^+\to D_s^+D_s^-K^+$ Decay,''
Phys. Rev. Lett. \textbf{131} (2023) no.7, 071901
%doi:10.1103/PhysRevLett.131.071901
%[arXiv:2210.15153 [hep-ex]].
%43 citations counted in INSPIRE as of 03 Jul 2024

%\cite{Bayar:2022dqa}
\bibitem{Bayar:2022dqa}
M.~Bayar, A.~Feijoo and E.~Oset,
``X(3960) seen in $D_s^+D_s^-$ as the X(3930) state seen in $D^+D^-$,''
Phys. Rev. D \textbf{107} (2023) no.3, 034007
%doi:10.1103/PhysRevD.107.034007
%[arXiv:2207.08490 [hep-ph]].
%22 citations counted in INSPIRE as of 26 Jun 2024

%\cite{Chen:2023eix}
\bibitem{Chen:2023eix}
Y.~Chen, H.~Chen, C.~Meng, H.~R.~Qi and H.~Q.~Zheng,
``On the nature of X(3960),''
Eur. Phys. J. C \textbf{83} (2023) no.5, 381
%doi:10.1140/epjc/s10052-023-11527-6
%[arXiv:2302.06278 [hep-ph]].
%8 citations counted in INSPIRE as of 02 Jul 2024

%\cite{Ding:2023yuo}
\bibitem{Ding:2023yuo}
Z.~m.~Ding and J.~He,
``Combined analysis on nature of $X(3960)$, $\chi_{c0}(3930)$, and $X_0(4140)$,''
Eur. Phys. J. C \textbf{83} (2023) no.9, 806
%doi:10.1140/epjc/s10052-023-11977-y
%[arXiv:2308.14264 [hep-ph]].
%2 citations counted in INSPIRE as of 26 Jun 2024

%\cite{Braaten:2004fk}
\bibitem{Braaten:2004fk}
E.~Braaten, M.~Kusunoki and S.~Nussinov,
%``Production of the X(3870) in B meson decay by the coalescence of charm mesons,''
Phys. Rev. Lett. \textbf{93} (2004), 162001
%doi:10.1103/PhysRevLett.93.162001
%[arXiv:hep-ph/0404161 [hep-ph]].
%79 citations counted in INSPIRE as of 26 Jun 2024

%\cite{LHCb:2022dvn}
\bibitem{LHCb:2022dvn}
R.~Aaij \textit{et al.} [LHCb],
``First observation of the $B^+\to D_s^+D_s^-K^+$ decay,''
Phys. Rev. D \textbf{108} (2023), 034012
%doi:10.1103/PhysRevD.108.034012
%[arXiv:2211.05034 [hep-ex]].
%12 citations counted in INSPIRE as of 26 Jun 2024

%\cite{Bando:1984ej}
\bibitem{Bando:1984ej}
M.~Bando, T.~Kugo, S.~Uehara, K.~Yamawaki and T.~Yanagida,
``Is rho Meson a Dynamical Gauge Boson of Hidden Local Symmetry?,''
Phys. Rev. Lett. \textbf{54} (1985), 1215
%doi:10.1103/PhysRevLett.54.1215
%1013 citations counted in INSPIRE as of 26 Jun 2024

%\cite{Bando:1987br}
\bibitem{Bando:1987br}
M.~Bando, T.~Kugo and K.~Yamawaki,
``Nonlinear Realization and Hidden Local Symmetries,''
Phys. Rept. \textbf{164} (1988), 217-314
%doi:10.1016/0370-1573(88)90019-1
%1400 citations counted in INSPIRE as of 26 Jun 2024

%\cite{Nagahiro:2008cv}
\bibitem{Nagahiro:2008cv}
H.~Nagahiro, L.~Roca, A.~Hosaka and E.~Oset,
``Hidden gauge formalism for the radiative decays of axial-vector mesons,''
Phys. Rev. D \textbf{79} (2009), 014015
%doi:10.1103/PhysRevD.79.014015
%[arXiv:0809.0943 [hep-ph]].
%145 citations counted in INSPIRE as of 26 Jun 2024

%\cite{Cheng:1992xi}
\bibitem{Cheng:1992xi}
H.~Y.~Cheng, C.~Y.~Cheung, G.~L.~Lin, Y.~C.~Lin, T.~M.~Yan and H.~L.~Yu,
``Chiral Lagrangians for radiative decays of heavy hadrons,''
Phys. Rev. D \textbf{47} (1993), 1030-1042
%doi:10.1103/PhysRevD.47.1030
%[arXiv:hep-ph/9209262 [hep-ph]].
%181 citations counted in INSPIRE as of 26 Jun 2024

%\cite{Yan:1992gz}
\bibitem{Yan:1992gz}
T.~M.~Yan, H.~Y.~Cheng, C.~Y.~Cheung, G.~L.~Lin, Y.~C.~Lin and H.~L.~Yu,
``Heavy quark symmetry and chiral dynamics,''
Phys. Rev. D \textbf{46} (1992), 1148-1164
[erratum: Phys. Rev. D \textbf{55} (1997), 5851]
%doi:10.1103/PhysRevD.46.1148
%782 citations counted in INSPIRE as of 26 Jun 2024

%\cite{Wise:1992hn}
\bibitem{Wise:1992hn}
M.~B.~Wise,
``Chiral perturbation theory for hadrons containing a heavy quark,''
Phys. Rev. D \textbf{45} (1992) no.7, R2188
%doi:10.1103/PhysRevD.45.R2188
%884 citations counted in INSPIRE as of 26 Jun 2024

%\cite{Burdman:1992gh}
\bibitem{Burdman:1992gh}
G.~Burdman and J.~F.~Donoghue,
``Union of chiral and heavy quark symmetries,''
Phys. Lett. B \textbf{280} (1992), 287-291
%doi:10.1016/0370-2693(92)90068-F
%688 citations counted in INSPIRE as of 26 Jun 2024

%\cite{Casalbuoni:1996pg}
\bibitem{Casalbuoni:1996pg}
R.~Casalbuoni, A.~Deandrea, N.~Di Bartolomeo, R.~Gatto, F.~Feruglio and G.~Nardulli,
``Phenomenology of heavy meson chiral Lagrangians,''
Phys. Rept. \textbf{281} (1997), 145-238
%doi:10.1016/S0370-1573(96)00027-0
%[arXiv:hep-ph/9605342 [hep-ph]].
%700 citations counted in INSPIRE as of 26 Jun 2024

%\cite{Falk:1992cx}
\bibitem{Falk:1992cx}
A.~F.~Falk and M.~E.~Luke,
``Strong decays of excited heavy mesons in chiral perturbation theory,''
Phys. Lett. B \textbf{292} (1992), 119-127
%doi:10.1016/0370-2693(92)90618-E
%[arXiv:hep-ph/9206241 [hep-ph]].
%271 citations counted in INSPIRE as of 26 Jun 2024

%\cite{Isola:2003fh}
\bibitem{Isola:2003fh}
C.~Isola, M.~Ladisa, G.~Nardulli and P.~Santorelli,
``Charming penguins in $B\to K^* \pi, K(\rho, \omega, \phi)$ decays,''
Phys. Rev. D \textbf{68} (2003), 114001
%doi:10.1103/PhysRevD.68.114001
%[arXiv:hep-ph/0307367 [hep-ph]].
%183 citations counted in INSPIRE as of 26 Jun 2024

%\cite{Liu:2009qhy}
\bibitem{Liu:2009qhy}
X.~Liu, Z.~G.~Luo, Y.~R.~Liu and S.~L.~Zhu,
``X(3872) and Other Possible Heavy Molecular States,''
Eur. Phys. J. C \textbf{61} (2009), 411-428
%doi:10.1140/epjc/s10052-009-1020-4
%[arXiv:0808.0073 [hep-ph]].
%230 citations counted in INSPIRE as of 26 Jun 2024


%\cite{Chen:2019asm}
\bibitem{Chen:2019asm}
R.~Chen, Z.~F.~Sun, X.~Liu and S.~L.~Zhu,
``Strong LHCb evidence supporting the existence of the hidden-charm molecular pentaquarks,''
Phys. Rev. D \textbf{100} (2019) no.1, 011502
%doi:10.1103/PhysRevD.100.011502
%[arXiv:1903.11013 [hep-ph]].
%194 citations counted in INSPIRE as of 26 Jun 2024

%\cite{Oh:2000qr}
\bibitem{Oh:2000qr}
Y.~s.~Oh, T.~Song and S.~H.~Lee,
``$J/\psi$ absorption by $\pi$ and $\rho$ mesons in meson exchange model with anomalous parity interactions,''
Phys. Rev. C \textbf{63} (2001), 034901
%doi:10.1103/PhysRevC.63.034901
%[arXiv:nucl-th/0010064 [nucl-th]].
%217 citations counted in INSPIRE as of 26 Jun 2024

%\cite{He:2015mja}
\bibitem{He:2015mja}
J.~He,
``The $Z_c(3900)$ as a resonance from the $D\bar{D}^*$ interaction,''
Phys. Rev. D \textbf{92} (2015) no.3, 034004
%doi:10.1103/PhysRevD.92.034004
%[arXiv:1505.05379 [hep-ph]].
%67 citations counted in INSPIRE as of 26 Jun 2024



%\cite{Gross:2008ps}
\bibitem{Gross:2008ps}
F.~Gross and A.~Stadler,
``Covariant spectator theory of np scattering: Phase shifts obtained from precision fits to data below 350-MeV,''
Phys. Rev. C \textbf{78} (2008), 014005
%doi:10.1103/PhysRevC.78.014005
%[arXiv:0802.1552 [nucl-th]].
%148 citations counted in INSPIRE as of 11 Jul 2024



%\cite{He:2014nya}
\bibitem{He:2014nya}
J.~He,
``Study of the $B\bar{B}^*/D\bar{D}^*$ bound states in a Bethe-Salpeter approach,''
Phys. Rev. D \textbf{90} (2014) no.7, 076008
%doi:10.1103/PhysRevD.90.076008
%[arXiv:1409.8506 [hep-ph]].
%54 citations counted in INSPIRE as of 26 Jun 2024

%\cite{He:2017lhy}
\bibitem{He:2017lhy}
J.~He and D.~Y.~Chen,
``$Z_c(3900)/Z_c(3885)$ as a virtual state from $\pi J/\psi-\bar{D}^*D$ interaction,''
Eur. Phys. J. C \textbf{78} (2018) no.2, 94
%doi:10.1140/epjc/s10052-018-5580-z
%[arXiv:1712.05653 [hep-ph]].
%27 citations counted in INSPIRE as of 26 Jun 2024

%\cite{He:2015yva}
\bibitem{He:2015yva}
J.~He,
``Internal structures of the nucleon resonances $N(1875)$ and $N(2120)$,''
Phys. Rev. C \textbf{91} (2015) no.1, 018201
%doi:10.1103/PhysRevC.91.018201
%[arXiv:1501.00522 [nucl-th]].
%36 citations counted in INSPIRE as of 26 Jun 2024

%\cite{He:2017aps}
\bibitem{He:2017aps}
J.~He,
``Nucleon resonances $N(1875)$ and $N(2100)$ as strange partners of LHCb pentaquarks,''
Phys. Rev. D \textbf{95} (2017) no.7, 074031
%doi:10.1103/PhysRevD.95.074031
%[arXiv:1701.03738 [hep-ph]].
%47 citations counted in INSPIRE as of 26 Jun 2024


%\cite{Blatt:1952ije}
\bibitem{Blatt:1952ije}
J.~M.~Blatt and V.~F.~Weisskopf,
``Theoretical nuclear physics,''
Springer, 1952,
ISBN 978-0-471-08019-0
%doi:10.1007/978-1-4612-9959-2
%401 citations counted in INSPIRE as of 26 Jun 2024

%\cite{BaBar:2010wqe}
\bibitem{BaBar:2010wqe}
P.~del Amo Sanchez \textit{et al.} [BaBar],
``Dalitz plot analysis of $D_s^+ \to K^+ K^- \pi^+$,''
Phys. Rev. D \textbf{83} (2011), 052001
%doi:10.1103/PhysRevD.83.052001
%[arXiv:1011.4190 [hep-ex]].
%81 citations counted in INSPIRE as of 26 Jun 2024



\end{thebibliography}

\end{document}